\documentstyle[aaspp4,11pt,psfig,sidef]{article}

\received{5 October 1998}

\lefthead{Padgett et al.}
\righthead{HST/NICMOS Imaging of Disks and Envelopes Around Very Young Stars}

\begin{document}

\title{HST/NICMOS Imaging of Disks and Envelopes Around Very Young 
Stars\footnote{Based on observations with the NASA/ESA Hubble 
Space Telescope obtained at the Space Telescope Science Institute, which is 
operated by the Association of Universities for Research in Astronomy, Inc., 
under NASA contract NAS5-26555.}}

\author{Deborah L.\ Padgett\altaffilmark{2}, Wolfgang Brandner\altaffilmark{2}, 
Karl R.\ Stapelfeldt\altaffilmark{3},\\ 
Stephen E. Strom\altaffilmark{4}, Susan Terebey\altaffilmark{5},
David Koerner\altaffilmark{6}}
\affil{$^2$Caltech - JPL/IPAC, Mail Code 100-22, Pasadena, CA 91125, USA}
\authoremail{dlp@ipac.caltech.edu}
\affil{$^3$JPL, 4800 Oak Grove Drive, Mail Stop 183-900, Pasadena, CA 
91109, USA}
\affil{$^4$NOAO, 950 N. Cherry Ave., Tucson, AZ 85726}
\affil{$^5$Extrasolar Research Corporation, 720 Magnolia Ave., Pasadena, CA
91106, USA}
\affil{$^6$University of Pennsylvania, Dept. of Physics and Astronomy,
209 South 33rd Street, Philadelphia, PA 19104-6396, USA}

\begin{abstract} 

We present HST/NICMOS observations with $\sim$ 0$\farcs$1
$\approx$ 15 AU resolution of six young stellar objects in the Taurus 
star-formation region. The targets of our survey are three Class I IRAS sources 
(IRAS 04016+2610, IRAS 04248+2612, and IRAS 04302+2247) and three 
low-luminosity stars (DG Tau B, Haro 6-5B, and CoKu Tau/1) 
associated with Herbig Haro jets. The broad-band images show that 
the near-infrared radiation from these sources is dominated 
by light scattered from dusty circumstellar material 
distributed in a region 10 - 15 times the size of our solar 
system. Although the detailed morphologies of the individual 
objects are unique, the observed young stellar objects share 
common features. All of the circumstellar reflection nebulae are 
crossed by dark lanes from 500 - 900 AU in extent and from 
less than 50 to 350 AU in apparent thickness. The absorption 
lanes extend perpendicular to known optical and millimeter 
outflows in these sources. We interpret the dark lanes as 
optically thick circumstellar disks seen in silhouette 
against bright reflection nebulosity. The bipolar reflection 
nebulae extending perpendicular to the dust lanes appear to 
be produced by scattering from the upper and lower surfaces 
of the disks and from dusty material within or on the walls 
of the outflow cavities. Out of five objects in which the 
central source is directly detected, two are found to be 
subarcsecond binaries. This mini-survey is the highest
resolution near-infrared study to date of circumstellar
environments around solar-type stars with age $\leq$ 1 Myr. 
\end{abstract}


\section{Introduction}

Theoretical and observational investigations of star formation
suggest that the formation process for low mass stars 
results naturally in the formation of a 
circumstellar disk which may then evolve into a planetary system. The 
current paradigm for the birth of single stars
(Adams, Lada, \& Shu 1987; Strom 1994, etc.) includes the following phases:

$\bullet$ 
Within a dusty molecular infall envelope, a star + nebular disk form. 
Most of the material destined for the star falls first onto the disk, 
due to the non-zero angular momentum of the infalling gas.
As gas accretes onto the star, an accretion-driven stellar or disk wind 
develops which begins to clear envelope gas away from the rotational poles 
of the system. This is the ``embedded young stellar object" (YSO) phase with a 
``Class I'' spectral energy distribution (SED) characterized by a
spectrum which rises beyond 2 $\micron$ because almost all the light
from the star is absorbed and re-radiated by circumstellar dust at long
wavelengths. In the most embedded objects, most of the scattered 
light may come from walls of the outflow cavity in the envelope,
giving a ``cometary'' appearance to the circumstellar reflection nebulosity.
Some extreme ``Class 0'' objects are completely invisible at wavelengths 
shorter than 25 microns and probably represent the early stages of the
embedded YSO phase (Andr\'e, Ward-Thompson, \& Barsony 1993). 
This phase is thought to last for a few times 10$^5$ yr.

$\bullet$ The infalling envelope disperses, leaving an optically
visible ``classical'' T Tauri star (CTTS) with a circumstellar disk.
Accretion through the disk slows, but continues as the star gains the final few 
percent of its mass. 
This ``Class II'' phase lasts from 10$^6$ to
10$^7$ yr and is characterized by a spectrum which has near-to-far IR
emission in excess of photospheric values, but is flat or falling
longward of 2 $\micron$.  Long wavelength emission is diminished since the 
solid angle subtended by the disk is smaller than the envelope, intercepting 
and reprocessing less of the radiation from accretion processes and the 
stellar photosphere.  As the envelope disperses, light reflected from the top 
and/or bottom surfaces of an optically thick disk will dominate the nebulosity, 
which will appear more ``flattened'' than for Class I sources.

$\bullet$ Accretion through the disk subsides due to lack of replenishment
or disk gaps created by planet formation. The disk becomes optically thin, 
and the system evolves into a ``weak-line'' T Tauri star (WTTS). Its
``Class III'' spectrum is basically that of a stellar photosphere with
a small amount of infrared excess at mid and far IR wavelengths if
an optically thin protoplanetary or planetary debris disk remains. 
These sources would look like the optically thin disk around Beta Pictoris 
(Backman $\&$ Paresce 1993) if edge-on, with starlight dominating a thin, 
bright, elongated nebula. Unfortunately, since the IR emission from these disks
are below the IRAS detection limits for nearby star-formation regions, there
are currently only loose constraints on the timescale for dispersal of optically
thin or ``debris'' disks.

\par
This picture provides a logical context for understanding the ensemble
of infrared SEDs observed for young stars (Adams, Lada, $\&$ Shu  1987),
as well as the sequence of millimeter continuum brightnesses among
YSOs (Saraceno et al. 1996; Andr\'e, Ward-Thompson, \& Barsony 1993).
However, there is as yet little direct observational confirmation at spatial 
scales which distinguish the morphology of the circumstellar material.
In particular, empirical information is lacking re:
(i) large-scale geometry and distribution of material in infalling envelopes; 
(ii) the geometry and extent of wind-created cavities; (iii) the large-scale
distribution of material within the disks. Hubble Space Telescope (HST)
optical (WFPC2) and near-infrared (NICMOS) observations potentially enable 
us for the first time to examine the morphology of circumstellar environments 
during all evolutionary stages with spatial resolution of 10 - 30 AU.
At this resolution, we can readily resolve and quantify structure in
envelopes (expected size $\sim$ 1000 AU) and infer information
on the large scale structure of disks (which are expected
to have sizes spanning the range 10 to several hundred AU).
Our group is carrying out a series of imaging programs
aimed at quantifying the morphology of YSOs spanning the full
range of evolutionary states. Our goal is to characterize the basic 
morphologies of objects in each SED class observable by HST, 
to identify common features
among objects of each class, to summarize complementary ground-based
data, and to identify those features which correspond to current
YSO models as well as those that are not/cannot yet be predicted
from such models.

\par

Ground-based observations have revealed that many Class I young stellar objects 
are nebulous in the optical and near infrared. Tamura et al. (1991) 
found that Class I objects which excite outflows are almost always associated
with compact and/or extended near-infrared nebulae. The observations of
Kenyon et al. (1993) revealed that Class I YSOs in the Taurus star-formation
region (distance 140 pc; Kenyon, Dobrzycka, \& Hartmann 1994)
often display a NIR cometary morphology which can be fit by models
of a flattened envelope with a polar outflow cavity. Many of the
Taurus Class I sources are also highly polarized, indicating the
dominance of scattered light at short wavelengths in these systems
(Whitney, Kenyon, $\&$ Gomez 1997).  Lucas \& Roche (1997) were able 
to characterize the structure of several Taurus Class I sources on 
subarcsecond scales and found that these sources were completely nebulous at 
1 $\micron$. A recent study by Kenyon et al. (1998) detected about half of the 
Taurus Class I sources in an optical spectroscopic survey using the Palomar 
5-meter and the MMT, during which they determined spectral types for several of 
the central sources.

\par
Previous HST observations of young stellar objects have demonstrated 
that high angular resolution imaging of circumstellar morphology
can make significant contributions to the study of young stellar 
object disks and envelopes.  HH 30 has been revealed to be a textbook 
validation of the circumstellar disk/wind paradigm (Burrows et al.\ 1996),
with bipolar reflection nebulae separated by an optically thick flared 
absorption disk 450 AU in diameter and an extremely narrow jet emanating
from within 30 AU of the central star.  HST/WFPC2 imaging has also shown that 
HL Tauri, previously thought to be a directly visible classical T Tauri star,
is actually an embedded system seen entirely by scattered light at optical 
wavelengths (Stapelfeldt et al.\ 1995a).

Our target sample consists of six established and probable Class I young
stellar objects. The objects include three low luminosity IRAS point sources 
(IRAS 04016+2610, IRAS 04248+2612, IRAS 04302+2247) and three Herbig-Haro jet 
exciting stars (DG Tau B, Haro 6-5B, CoKu Tau/1). 
The three IRAS stars are Class I infrared sources. 
The SED classification of the remaining sources are uncertain because they
are confused with bright nearby YSOs in the large IRAS beams.
These objects span a considerable range in millimeter continuum flux,
which is listed in Table 1. Millimeter continuum flux is 
interpreted as optically thin blackbody emission from the
dust around young stellar objects, and the amount of flux
is related to the amount of circumstellar material (Beckwith
et al. 1990; Osterloh $\&$ Beckwith 1995). The span
of millimeter continuum fluxes for our sample objects corresponds
to more than an order of magnitude in circumstellar masses, from 
about 10$^{-2}$ M$_{\odot}$ for F$_{1 mm}$ $\sim$ 200 mJy to
$\leq$ 10$^{-3}$ M$_{\odot}$ for 10 mJy (Beckwith et al. 1990).
Of the six objects in our sample, four have unresolved 
continuum peaks at 1.3 mm as mapped by millimeter interferometry.
indicating the presence of an unresolved dust disk (IRAS 04302+2247,
Padgett et al. 1999; DG Tau B, Stapelfeldt et al. 1999;
Haro 6-5B, Dutrey et al. 1996; IRAS 04016+2610, Hogerheijde
et al. 1997). The two remaining YSOs (CoKu Tau/1,
IRAS 04248+2612) have only been mapped at millimeter wavelengths
with relatively large (12\farcs) beams; however, they
have the lowest circumstellar masses in any case. 
In order to facilitate modelling at multiple wavelengths,
we chose objects for which HST/WFPC2 observations were either
planned or already taken.  The pre-existing WFPC2 data
enabled us to select several objects with known
close to edge-on ``disk'' orientations.
As predicted by Bastien \& Menard (1990) and Whitney \& Hartmann (1992),
and observationally demonstrated by 
Burrows et al.\ (1996), the visibility of optically thick
circumstellar disks is maximized when the disk material itself occults the star,
improving the contrast of the nebulosity
relative to the stellar point source. The low luminosity jet
source CoKu Tau/1 was chosen on the basis of its faintness at 2 microns
(Stapelfeldt et al.\ 1997b) as well as the low radial velocity of its
jet, indicating that its optical outflow is nearly in the plane of the
sky (Eisl\"offel \& Mundt 1998).
Detailed comparison
and multiwavelength modeling
of HST/WFPC2 and HST/NICMOS images of some of these YSOs
will be presented in future papers. The current paper 
attempts to characterize the morphology of the observed Taurus
YSOs at a scale of tens of AU at wavelengths from 1.1 to 2 $\micron$.

%
%

\section{Observations and Data Analysis} 

\subsection{HST/NICMOS observations}

The observations have been obtained with HST/NICMOS (see Thompson et al.\
1998 for a description of the NICMOS instrument) between August and 
December 1997.  We used NIC2 and the filters F110W, F160W, F187W, and F205W. 
Exposure times in individual bands are shown in Table 2. Each target was 
imaged twice in each band on different portions of the NICMOS array, using the 
predefined spiral dither pattern. The two-step dither pattern allowed us to 
compensate for bad pixel/columns and the area blocked by the coronographic mask.

%
%

\subsection{Re-reduction of NICMOS data}

The raw data were re-reduced using the IRAF/STSDAS package CALNICA V3.1 and
the latest set of reference files. Pixel masks based on the template
bad pixel masks made available through STScI were created for each
individual image and edited to include the actual position
of the coronographic spot, transient bad pixels, and cosmic ray events
which had not been detected and cleared out by CALNICA. The ``photometrically
challenged column \#128'' (Bergeron \& Skinner 1997) was masked as well.

The offsets between the two subsets of each association were computed
by cross-correlating the individual subsets. Using the IRAF/STSDAS package 
DRIZZLE (Fruchter \& Hook 1997), the subsets were then re-mapped into one frame
with the same pixel scale, but taking into account subpixel shifts between the
individual frames, and the information in the bad pixel masks. 

\subsection{Deconvolution of images}

The NICMOS-PSF, although well defined, has very extended wings. In order
to improve the contrast of our images, we decided to deconvolve them
with the aim to remove (or at least attenuate) the extended PSF features.
After remapping, the images were rebinned by a factor of 2 in X and Y,
and deconvolved with a synthetic point-spread function (PSF). The PSFs were
computed with TINYTIM V4.4 (Krist 1997) and oversampled by a factor of 2.
Because of finite sampling, deconvolution of point sources superposed on a 
diffuse bright background in general leads to ``ringing'' artifacts, unless 
one uses a narrower PSF (Lucy 1990, Snyder 1990). We therefore first 
deconvolved the TINYTIM PSF itself with Gaussian. The FWHM of the Gaussian
was chosen to match the actual sampling of our data.

The deconvolution of the data was done within IDL, using an iterative Poisson 
maximum
likelihood algorithm based on the Richardson-Lucy-scheme (Richardson 1972,
Lucy 1974) and described in detail by Blecha \& Richard (1989). The iteration 
was stopped after 5 iterations in F110W, 8-10 iterations in F160W, 10-15
iterations in F187W, and 15-20 iterations in F205W.
The deconvolution reduced the flux in the extended PSF features
by about a factor of 3, and gave the deconvolved images in the various filters
approximately the same spatial resolution.

\subsection{SYNPHOT flux calibration}

In order to get an estimate on the brightness of our sources, we calculated
the count rates within an aperture of 3\farcs8 radius (50 pixel). 
Sky/background values were measured on areas of the NICMOS frames which were 
free of diffuse emission. 

Flux calibration was computed in an iterative approach, using the IRAF/STSDAS
package SYNPHOT. This step was necessary, as the photometric header keywords
provided by the NICMOS calibration pipeline assume a spectral energy 
distribution with a constant flux per unit wavelength. The spectral energy 
distribution of our sources, in particular those of the three class I sources
(IRAS 04016+2610, IRAS 04248+2612, and IRAS 04302+2247),
deviates significantly from this assumption. Using IDL and IRAF, we first 
computed a model spectral energy distribution, based on the photometric header 
keywords. We then ran SYNPHOT for the model SED and computed a new set of 
photometric conversion factors. This allowed us to create a revised model SED 
and to derive improved photometric conversion factors. The iterative procedure 
was repeated until the flux values in the individual filters changed by less 
than 2\% in consecutive iterations.

Depending on the actual shape of the spectral energy distribution, the 
resulting flux values in individual filters are between 25\% and 180\% of the
initial estimates as derived from the pipeline header keywords. Aperture
photometry of the sources is reported in Table 3. The quoted uncertainties
are statistical errors. Systematic errors in the flux calibration
(due to the unknown shape of the spectral energy distribution below 1$\mu$m
and above 2$\mu$m) amount to an additional uncertainty of 10\%--15\%.

%
%

\subsection{Photometry of Point Sources and Binaries}

Magnitudes of point sources were determined using a 0\farcs5 aperture,
which excludes most of the extended, diffuse emission. The separation,
position angle, and brightness ratio of the close binary sources were
determined by least squares fitting based on Gaussian PSFs (see Brandner et
al.\ 1996 for details). Point source photometry is shown in Table 4.
The quoted uncertainties are statistical errors. For the binary companions
deviations of the actual PSF from a Gaussian result in an additional
uncertainty of $\approx$10\%.

%
%
%
%

\section{Results for Individual YSOs}

Figure 1(a) - (f) show a three-color composite image
for each of the young stellar objects in our sample.
In this figure, we present ``pseudo-true color'' (sensitivity normalized) 
mappings of F110W to blue, F160W to green, and F205W to red. The scale and 
orientation of each image are depicted in the Figures. 
Figures 2(a) - (f) are the flux calibrated contour maps of each
source for the F160W images. Although
the images have been deconvolved and reconvolved to a common resolution,
residual PSF features remain, especially in the F205W images. These artifacts
appear as multicolored spots and rings surrounding the stellar PSF.
We are therefore wary of interpreting structure near the Airy rings
of the stars.
The deconvolution has also added speckle noise to the background
which has been de-emphasized in the stretches presented here.

\par
   The results are presented in order of decreasing millimeter
continuum flux. To the extent that the objects have comparable 
luminosities, this is equivalent to sorting from largest 
to smallest circumstellar masses. 

\subsection{DG Tau B}

DG Tau B is a low-luminosity jet source located 1$'$ southwest
of DG Tau, and its IR SED is confused with this brighter optical source.
DG Tau B was first identified as a Herbig-Haro jet exciting star
by Mundt \& Fried (1983), and the kinematics of its optical outflow
(HH 159) has been the subject of several papers (Eisl\"offel \& Mundt 1998,
Mundt, Brugel, \& B\"uhrke 1987). Radial velocities indicate that the
jet is directed to within 15$^{\circ}$ of the plane of the sky. 
DG Tau B also drives a large molecular jet which
is spatially coincident with its redshifted optical jet
(Mitchell et al.\ 1997, 1994). The DG Tau B source has been detected 
in 6 cm radio continuum emission (Rodriguez et al. 1995) 
which is further evidence of this system's youth. Stapelfeldt et al.
(1997) found that DG Tau B is resolved in HST/WFPC2 images as a compact
bipolar nebula with no optically visible star.

   In the HST/NICMOS composite image presented in Figure 1a, DG Tau B appears
as a bipolar reflection nebula. The eastern lobe appears V-shaped, with
an axis of symmetry coinciding with the direction of the blueshifted jet 
(Eisl\"offel \& Mundt 1998). Therefore, we conclude that the ``V'' traces
the walls of the blueshifted outflow cavity.  At 2 $\micron$, the stellar PSF 
becomes visible at the apex of the ``V''. Immediately to the southwest of the 
central source is a bright spot in the nebula along the inner edge of ``V''. 
Although this feature might be due to a companion source, the spot itself is
extended relative to the PSF and, therefore, cannot be a companion
seen directly. There appear to be no significant color gradients across
the cavity walls.
The western lobe of the nebula is several times fainter
and has a much narrower opening angle. It encompasses the redshifted
optical and CO jets. Several knots in the redshifted jet are visible
in the 1 $\micron$ (F110W) image (Figure 3), probably due to  
1.26 $\micron$ [Fe II] emission. Knots in the blueshifted jet are
not obvious against the eastern lobe's bright reflection nebulosity. 
Separating the two lobes of bright nebulosity is a thick dust lane
perpendicular to the jet which can be traced to a length of 600 AU.  A dense 
bar of $^{13}$CO(1-0) emission has recently been detected at this location
by Stapelfeldt et al. (1999) using the Owens Valley Radio Observatory
(OVRO) Millimeter Array. 

\subsection{IRAS 04016+2610}

IRAS 04016+2610 is a Class I infrared source with L$_{bol}$ = 3.7 L$_{\odot}$
located in the L1489 dark cloud of the Taurus star-forming region (Kenyon
\& Hartmann 1995). Millimeter-wave observations of this source have detected
an extended molecular gas core (Hogerheijde et al.\
1998, Ohashi et al.\ 1991). This IRAS source
appears as a scattered light nebula in the optical and near infrared
(Lucas \& Roche 1997, Whitney et al.\ 1997), but is centrally condensed
at 2 microns. IRAS 04016+2610 powers a low velocity molecular outflow 
adjacent to the source (Hogerheijde et al.\ 1998) and has been
suggested as the powering source of a more extended molecular outflow
to the northwest (Moriarty-Schieven et al.\ 1992, Terebey et al.\ 1989).
Several Herbig-Haro objects coincide with the lobes of these
outflows (Gomez et al.\ 1997). 

In Figure 1b, IRAS 04016+2610 appears as a unipolar reflection nebula 
with a point source at the apex at 1.6 $\micron$ and 2 $\micron$. The
multicolored specks around the point source are PSF artifacts which were
not entirely removed by the deconvolution process.  Among the sources in our 
NICMOS imaging sample, IRAS 04016+2610 is the brightest point source at 2 
$\micron$ (Table 2), with about 75$\%$ of the flux concentrated in the PSF 
(Table 3). In the 1 $\micron$ image, the object is dominated by scattered light,
and the highest surface brightness region is adjacent to the point
source position seen at longer wavelengths. This bright patch
is separated from the main part of the reflection nebula by an 
area of extinction extended roughly east-west.  This feature is
visible as a swath of reddening which cuts diagonally across the
reflection nebula in Figure 1b. Between 1 $\micron$ and 1.6 $\micron$, the
shape of the reflection nebula changes significantly, from a V-shaped
morphology at 1 $\micron$ to a broader bowl-shaped nebula at 
longer wavelengths.  The symmetry axis of this nebula is well aligned
with adjacent optical and blueshifted millimeter outflows (Hogerheijde et al. 
1998, Gomez et al. 1997). About 3$''$ NNE of the point source is a faint
triangular patch of reflection nebula detached from the main nebula. 
A dark lane oriented at PA$=$ 80$^{\circ}$ separates this counternebula from
the main nebula but does not extend to the other side of
main nebula's symmetry axis.  The dust lane can be traced for at least 600 
AU before it loses definition in the extremely dark region to the northwest of 
the star. The position angle of this dark lane matches the orientation of the 
elongated $^{13}$CO(1-0) and HCO$^+$ structures mapped by Hogerheijde et al. (1998).  

\subsection{IRAS 04302+2247}

IRAS 04302+2247 is a Class I source with estimated L$_{bol}$ = 0.34 L$_{\odot}$
(Kenyon \& Hartmann 1995) located in the vicinity of the Lynds 1536b dark
cloud. Undetected at 12 microns by IRAS, the infrared SED of this source peaks
around 100 microns with a flux of about 10 Jy (Beichman et al.\ 1992).  
Bontemps et al. (1996) mapped 
a small, low-velocity CO outflow nearby which they associate 
with this source, and Gomez et al. (1997) have found two Herbig-Haro objects 
several arcminutes northwest overlying the blueshifted lobe of the outflow. 
IRAS 04302+2247 was studied in the near-IR at UKIRT by Lucas \& Roche (1997, 
1998), who published the first description of the remarkable dust lane
and bipolar nebula of this system and presented near-IR polarimetry.

IRAS 04302+2247 is surely among the more spectacular
young stellar objects observed by the Hubble Space Telescope. 
The HST/NICMOS near-IR appearance of this object in Figure 1c is dominated 
by the totally opaque band extending 900 AU north/south which bisects 
the scattered light nebulosity.  No point source is detected in this
source at any of the observed wavelengths.  The apparent thickness of the 
extinction band decreases by about 30$\%$ from 1 $\micron$ to 2 $\micron$, 
which accounts for the reddening seen along its edges. 
At 2 $\micron$, the dust lane appears thicker at its ends than at its
center. Although this dark feature has relatively straight edges at 1 $\micron$,
at 2 $\micron$, the lane is thicker at the ends than in the middle by a
factor of two. Owens Valley Millimeter Array mapping of this source in 
$^{13}$CO(1-0) to be presented in Padgett {\it et al.} (1999) indicate that 
the dark lane coincides with a dense rotating disk of molecular gas.
The dust lane of IRAS 04302+2247 may therefore be a large optically
thick circumstellar disk seen precisely edge-on.

The scattered light nebula of IRAS 04302+2247 is dramatically bipolar
in morphology, with approximately equal brightness between the eastern
and western lobes. The eastern lobe is only a few percent brighter
than the western lobe at 1 $\micron$ and 1.6 $\micron$, but is 15 $\%$
brighter at 2 $\micron$.  The shape of the nebular lobes is
roughly similar to the wings of a butterfly, giving the object its alias 
``Butterfly Star in Taurus'' (Lucas \& Roche 1997). At 1 $\micron$,
the brightest parts of the nebula are confined to the central region
along the presumed outflow axis perpendicular to the dust lane.
However, the nebular morphology changes
with increasing wavelength, from roughly ``V''-shaped at 1 $\micron$ to more 
flattened along the dust lane at 2 $\micron$.  Within each lobe are a variety of
bright filamentary structures subarcsecond in length and unresolved in
width, some of which are curved near the dust lane. There are also areas of 
non-uniform extinction within the lobes which suggest a rather clumpy
distribution of material. A prominent swath of extinction extends
asymmetrically across the northern part of the western reflection 
lobe, curving smoothly into the dust lane. The central region of each 
lobe is fainter than the outer edges. These extinction features divide
the bright lobes into quadrants, accounting for the ``quadrupolar'' morphology
seen by Lucas \& Roche (1997) and modeled as an opaque jet. One possibility
is that intervening clumps of absorbing material are superposed on the
symmetry axis of the reflection lobes. This interpretation would be
similar to one advocated by Stapelfeldt et al. (1995a) for HL Tau.
However, the darker zone does not appear to be edge-reddened as would be
expected for a dark clump. Another possibility is that the darker region 
is an evacuated zone along the presumed outflow axis which is cleared of the 
reflective streamers of material seen along the cavity walls. 

\subsection{Haro 6-5B}

Haro 6-5B, located about 20$''$ west of the FS Tauri T-Tauri
star binary, is the source of the HH 157 optical jet.
Because of its proximity to this brighter pre-main sequence system, the IRAS 
SED of Haro 6-5B is confused with FS Tau.  This YSO was first noted as a 
region of compact reflection nebulosity along the emission line jet (Mundt \& 
Fried 1983). The kinematics of the jet suggest that the outflow for this source 
is nearly in the plane of the sky (Eisl\"offel \& Mundt 1998). 
Millimeter interferometry indicates
that Haro 6-5B has compact $^{13}$CO(1-0) emission (Dutrey et al. 1996).
HST/WFPC2 imaging revealed that Haro 6-5B is a compact bipolar nebula bisected 
by a dust lane which is similar to models of a nearly edge-on optically thick 
disk (Krist et al. 1998). Therefore, Haro 6-5B appears very similar to the 
edge-on young stellar disk system HH 30 (Burrows et al.\ 1996).

HST/NICMOS images of Haro 6-5B generally resemble the HST/WFPC2 images
of Krist et al. (1998) with the exception that the point spread function
of the star is directly detected in the near-IR images. The NICMOS color
composite of Haro 6-5B is presented in Figure 1d. The source itself appears
as two parallel curved reflection nebulae separated by a dark lane 
about 600 AU in length. The bipolar nebula is most extended along the dark lane,
perpendicular to the optical outflow axis.  The northeastern reflection lobe, 
from which the blueshifted optical jet emerges, is brighter than its 
southwestern counterpart at all observed wavelengths. The PSF is visible 
in the 1.1 $\micron$ image at the base of the northeastern lobe and contributes 
60\% of the light at 2 $\micron$.  
Using the upper limit for stellar V magnitude from Krist et al. (1998) and
assuming a late type photosphere, we determine that the lower limit of 
extinction toward the Haro 6-5B star is A$_V$ $\approx$ 8. If an appreciable 
percentage of the emission identified as photospheric is actually produced by 
hot dust close to the star, the extinction could be larger. 
About 10$''$ north of Haro 6-5B is a large 
diffuse nebula detached from the circumstellar nebulosity.  This nebulosity was 
also detected by WFPC2 and given the designation R1 in Mundt, Ray, \& Raga 
(1991).  There is also a faint suggestion of at least one knot in the redshifted
jet in the F110W and F160W images.

\subsection{IRAS 04248+2612}

IRAS 04248+2612 is a Taurus Class I source with a luminosity of 0.36
L$_{\odot}$ (Kenyon \& Hartmann 1995). The object 
was weakly detected in HCO+ (Hogerheijde {\it et al.} 1997) by
the James Clerk Maxwell Telescope.  IRAS 04248+2612 apparently drives
a small molecular outflow (Moriarty-Schieven et al. 1992) which has
not been mapped. Also known as HH31 IRS, IRAS 04248+2612 is presumed to be
the exciting source for HH 31 and several other small Herbig-Haro objects (Gomez
et al. 1997). In the near infrared, this source has a complex 
bipolar reflection nebulosity which was studied with shift-and-add UKIRT 
infrared imaging and polarimetry by Lucas \& Roche (1997). Imaging
polarimetry performed by Whitney et al. (1997) led to their
suggestion that this source is seen close to edge-on. 

   In the NICMOS images, IRAS 04248+2612 appears as a long, curving
bipolar reflection nebula. The major axis of this nebula extends
for at least 10$''$ (1400 AU) north-south, bending significantly
east at the southern end and west at the northern end. The nearby string
of HH objects lie along the long axis of
the southern lobe close to the source, but curve eastward into
an S-shaped jet at greater distances, suggesting a time-varying
outflow axis (Gomez et al. 1997). The elongated reflection nebula
of IRAS 04248+2612 therefore appears to define outflow channels.
Although many YSO outflow cavities appear to be limb-brightened,
the reflection nebula in this system is centrally brightened.
Along the outflow axis within the southern lobe is an bright
elongate structure which appears helical. This ``corkscrew''
nebulosity extends about 420 AU southwards from the central source.
The northern lobe also has a bright structure which seems to
be the mirror image of the southern helix. However, it appears
to end or be disrupted within about 150 AU north of the central source.
Although the morphology of these structures suggests an outflow
origin, their similar brightness in all the wide NICMOS bands
seems to indicate that they are reflection rather than emission
nebulae.  An additional faint patch of reflection nebulosity is located about
750 AU to the northeast of the binary, detached from the northwest lobe
of reflection nebulosity. Polarization maps presented in Lucas $\&$ Roche (1997)
indicate that the position angle of the polarization vector is consistent
with illumination by the distant binary.

NICMOS imaging of the IRAS 04248+2612 central source reveals that it
is actually a close binary with projected separation
of $\sim$ 25 AU (Table 4). Although the components of the binary are 
comparable in brightness, the eastern star (A) appears slightly redder than 
its neighbor. The presence of a close binary in this system is in
accordance with its lower millimeter continuum flux relative to
other Class I sources in the study. In addition, orbital motion
of a binary jet source offers a plausible explanation for both
the helical dusty ``trail'' in the southern nebular lobe and the
large scale sinusoidal curving of the southern jet seen by Gomez et al.
(1997).  The central  binary appears to peek over the north edge of a dark
lane which pinches the bright nebulosity into bipolar
components. This apparent dust lane is at least 450 AU in diameter and appears
to extend along a position angle perpendicular to the outflow axis.
The PA of the absorption lane seems to be slightly offset from the
separation vector between the two stellar components. 

\subsection{CoKu Tau/1}

CoKu Tau/1 is another faint Herbig-Haro object exciting star located in
the L1495 cloud near the embedded Ae star Elias 3-1, which confuses
its 60 $\micron$ and 100 $\micron$ IRAS SED. CoKu Tau/1 is detected at
the shorter wavelength IRAS bands with F(12$\micron$) = 1.18 $\pm$ 0.26 Jy
and F(25$\micron$) = 2.74 $\pm$ 0.63 Jy (Weaver \& Jones 1992). Its
total luminosity is estimated at only 0.065 L$_{\odot}$ by Strom \& Strom 
(1994), who derived a spectral type of M2e. CoKu Tau/1 has been detected
in the radio continuum at 6 cm (Skinner, Brown, \& Stewart 1993), but
is undetected at 1.3 mm (see Table 1). This object is the source of the small
HH 156 bipolar jet (Strom et al. 1986); its kinematics suggest that the outflow
is near the plane of the sky (Eisl\"offel \& Mundt 1998).

In the HST/NICMOS near-IR images (Figure 1f), 
CoKu Tau/1 appears as a faint binary with four filamentary reflection
nebulae curving parabolically away from the central sources. 
Since the optical outflow is known to emerge from between
the southwestern ``horns'', we interpret these structures
as the limb-brightened walls of outflow cavities.  Within the northern cavity 
is a filamentary arc of material which forms a closed loop about 200 AU in size.
Although the morphology of this feature is suggestive of a dark clump backlit by
bright nebulosity, no enhanced reddening is seen within the loop.

Like IRAS 04248+2612, CoKu Tau/1 is a previously unrecognized binary
(see Table 4).  Both of these binary systems are
too faint at 2 microns to have been detected by published ground-based speckle
surveys (Ghez et al. 1992, Leinert et al. 1993, etc.). The CoKu Tau/1 
secondary is about a magnitude fainter than the primary and is somewhat redder. 
The filamentary loop in the northern outflow cavity is located 
in the vicinity of the secondary which suggests that it is a secondary
outflow cavity. 

HST/NICMOS images reveal a local minimum in surface brightness between the two
cavities and along the plane of the binary stars.  This feature is
suggestive of a dust lane which appears much thinner than the other dust lanes
seen in our survey. This might be a circumbinary ring or disk structure.
The mass of the structure would have to be small, since
the millimeter continuum limits the mass to less than 10$^{-3}$
M$_{\odot}$. However, very little dust mass is required to produce a disk 
which is optically thick at near-infrared wavelengths, and, therefore,
visible as a dust lane when at near edge-on inclinations.

%
%

\section{Discussion}

\subsection{Dust Lane Properties}

   All of the young stellar objects imaged in the current NICMOS survey
have a morphology which includes a dark lane crossing the scattered light
nebula. Table 5 lists morphological parameters for the dust lanes seen
in the NICMOS images of young stellar objects.
The lengths and thicknesses of dust lanes were determined by
making cuts across the feature midplane perpendicular
to the major axis. The minor axis was measured
by averaging 5 cuts through the center of the dark lane. 
In cases where the center of the dust lane is adjacent to a PSF,
we determined the dust lane thickness by taking the mean of cuts
on both sides of the PSF, beyond the Airy ring and other PSF artifacts.
The dip in the brightness profile caused by the dust lane was fit by a
Gaussian using the IRAF tool IMPLOT, and the derived 
full width half maximum of this feature is the ``apparent
thickness'' or minor axis given in the results section and Table 5.  
The major axes of the dark lanes were determined by noting the radial distance 
from the photocenter at which
the dip in the surface brightness profile was less than 10\% of its
maximum depth or where the feature widened to twice the FWHM at or near
the photocenter.  Although some of 
these features can be traced to greater distances, our intent was to place a 
lower limit on the disk extent without confusing it with the separation of 
cavity walls. Better quantification of disk parameters awaits the application
of multiple scattering models in future papers. ``PSF visibility''
gives the filter at which a stellar PSF was detected, and the ``+''
indicates that the PSF was detected at all longer wavelengths.

   The lengths of these dust lanes vary within our sample from 500 AU
to 1000 AU. The apparent thicknesses of these dark features range from 50 AU  
- 340 AU. These widths do not represent the actual scale height of dense 
material, but rather define surfaces where optical depth $\approx$ 1
at NIR wavelengths for dusty circumstellar 
structures. The apparent thicknesses of the dust lanes seem to be related to 
the amount of millimeter continuum, in that objects with more 1 mm emission 
have thicker lanes. In addition, comparison of the dust lane position angles 
(Table 5) with the position angles of known outflows (Table 1) reveals that the
lanes are perpendicular to outflows in almost every case. Finally,
in the dust lanes of IRAS 04016+2610, Haro 6-5B, and IRAS 04302+2247,
are spatially coincident with dense, possibly rotating,
molecular bars mapped by millimeter interferometry (Hogerheijde et al. 1998,
Dutrey et al. 1997, Padgett et al. 1999, Stapelfeldt et al. 1999).  

   Based on these lines of evidence,
we conclude that the absorption bands seen in the HST/NICMOS images
are probably optically thick circumstellar disks seen in silhouette against
reflection nebulosity. The same interpretation has been offered for
dark elongated features seen in HST imaging
of several other YSOs including HH 30 (Burrows et al.\ 1996)
Orion 114-426 (McCaughrean et al. 1998), 
Haro 6-5B (Krist et al. 1998), and HK Tau/c (Stapelfeldt et al. 1998). 
As explained in considerable detail by
these authors, optically thick disks are completely opaque at optical
and NIR wavelengths, with hundreds to thousands of magnitudes of
extinction on a line of sight through the midplane. Therefore,
the disks themselves are dark, while their upper and lower surfaces
are illuminated by the central source. Models for disks of this
sort are presented in Bastien \& Menard (1990), Whitney \& Hartmann (1992, 1993), Fischer,
Henning, \& Yorke (1996), Burrows et al.\ (1996), and Wood et al. (1998).
Flattened envelopes may also produce dust lanes, as the models of
Whitney \& Hartmann (1993) indicate; kinematics from millimeter
interferometry are required to differentiate these components
of circumstellar material.

   Scattered light models of optically thick circumstellar disks
and flattened envelopes predict that the appearance of a YSO will vary
according to disk mass and inclination relative to the line of sight.
Optically thick circumstellar disks viewed within a few degrees of
edge-on will entirely occult the star due to high extinction in
the midplane. Conversely, if they are viewed pole-on, the dynamic
range required to detect light reflected from the disk is predicted
by theory to be beyond the capability of early 1990's technology (Whitney et al.
1992).  Edge-on disk systems are detected only in scattered light even at 
mid-IR wavelengths (Sonnhalter et al. 1995). There are currently three known 
edge-on disk sources: HH 30, Orion 114-426, and HK Tau/c,
to which the current study can add a fourth - IRAS 04302+2247. Edge-on disks 
appear as concave nebulae of similar brightness which are separated by a 
flared dark lane.  The apparent thickness of the lane increases with disk mass, 
ranging from 0$\farcs$1 (at the distance of Taurus) for 10$^{-4}$ M$_{\odot}$ 
to ten times that for 10$^{-2}$ M$_{\odot}$ (Burrows et al. 1996).
In addition, the apparent thickness of the disk will
also vary with wavelength, becoming thinner at longer wavelengths.
In systems which are slightly more than 10$^{\circ}$ from edge-on,
the flared outer parts of the disk or the optically thick base of the
envelope may occult the star at short wavelengths, 
but a PSF may be visible at longer wavelengths. In addition, one nebula
will be considerably brighter than the other.  In the current sample,
DG Tau B fulfills these criteria, since the PSF is
effectively undetected at 1 $\micron$. We also note that no PSF was detected
for Haro 6-5B in WFPC2 8000\AA\ images (Krist et al. 1998); however,
the stellar PSF was detected for this source in all of our NIR bands. 
Unlike envelope sources, Class II T Tauri stars have no circumstellar material 
except in the disk plane; therefore, they rarely have the gradation of 
extinction seen in embedded YSOs (invisible at V, bright at K). Depending on 
their disk inclination, T Tauri stars will either be bright with little 
circumstellar extinction or nebulosity, or they will be almost invisible at 
optical and NIR wavelengths (Stapelfeldt et al. 1997).

  Although the distribution of circumstellar
material in YSOs is unquestionably flattened, the kinematics of this
material is still controversial. Models which postulate a flattened,
infalling envelope surrounding a small, geometrically flat rotating disk
are quite successful in explaining the scattered light distribution
around YSOs (Whitney \& Hartmann 1992, 1993; Hartmann et al. 1994).
Lucas \& Roche (1997) simulated the scattered light distribution in their
ground-based images of IRAS 04302+2247 with a model incorporating 
an Ulrich envelope. However, such models are less successful in
reproducing the scattered light distribution from HH 30. 
In this case, the circumstellar nebulosity is better modeled as a large
optically-thick edge-on circumstellar disk (Wood et al. 1998). 
Recent high-resolution molecular line observations are
beginning to clarify the kinematic nature of the material in these large 
``disks''.  Rotating CO structures around many YSOs with sizes from 100 AU - 
1600 AU have been mapped using millimeter interferometers (Koerner $\&$ Sargent 
1995, Dutrey et al. 1996, Hogerheidje et al. 1998). 
The disk around the T Tauri star GM Aurigae, which has been detected
in reflected light by HST (Stapelfeldt et al. 1995b), shows a Keplerian
rotation curve out to a radius of 500 AU (Dutrey et al. 1998).
Those YSOs in the current survey which have been imaged via millimeter
interferometry (IRAS 04016+2610, Hogerheijde et al. 1998; Haro 6-5B, 
Dutrey et al. 1996; DG Tau B, Stapelfeldt et al. 1999; IRAS 04302+2247,
Padgett et al. 1999) all show concentrations of dense molecular gas close to
the star. Unfortunately, the $\sim$ 3$\farcs$0 resolution at which these YSOs
have been mapped in the millimeter allows only a rough correspondence with 
features seen in HST images. OVRO Millimeter Array 
observation of IRAS 04302+2247 by Padgett et al. (1999) have confirmed that 
the dense material along the dust lane in this source also appears to be 
rotating. These observations suggest that the dense material within the YSO 
dark lanes seen by HST is rotating, and, possibly, centrifugally bound.

\subsection{Properties of Bipolar Nebulosity}

 In most of the YSOs in the current sample, the symmetry axis of the 
bright reflection nebulae (perpendicular to the dust lanes) 
correspond well to the outflow position angles
determined by previous studies. Table 1 lists outflow position angles, and Table
6 contains the position angle of the nebula symmetry axis ($\theta_{sa}$). 
The ``major axis'' in Table 6 is the extent of the nebula along the
symmetry axis, and the ``minor axis'' is the extent perpendicular
to the symmetry axis.  Only IRAS 04302+2247 is linked to a jet which has a 
very different position angle from the reflection nebula symmetry axis. 
HH 394 lies several arcminutes to the northwest of IRAS 04302+2247, forming a 
loose chain of HH objects which point back in the direction of the IRAS source
(Gomez et al. 1997). The HH objects coincide well with the position
of a blue-shifted clump of $^{12}$CO(1-0) which was presumed to be
part of the IRAS 04302+2247 molecular outflow by Bontemps et al. (1996).
However, this optical and molecular outflow has a position angle
which is only 20$^\circ$ offset from the dust lane (``disk'' plane)! One possible
explanation is that the disk is not perpendicular to the outflow
for this source. However, we prefer to conclude that these optical
and molecular outflows are probably unrelated to IRAS 04302+2247 and are 
instead part of an outflow from another, more distant source, as seen
frequently
in Orion (Reipurth et al. 1997) and the Perseus Molecular Cloud (Bally et al.
1997). A similar situation may exist for the extensive molecular
outflow located to the northwest of IRAS 04016+2610 (Terebey et al.
1989).

%
%

\par
The reflection nebulosity associated with the current sample of
young stellar objects is most often bipolar, although the detailed
structure of each nebula is unique. Since the axis of
known outflows tends to coincide with the symmetry axis of the
reflection nebulae and the lobes are often extended along the
outflow directions,  it appears that the bright nebulosity most
often traces the $\tau$ = 1 scattering surface of dusty material
in outflow cavities. Outflow cavities are presumed to represent the
polar regions of circumstellar disk/envelope systems which have been
cleared of dense gas by stellar jets (e.g. Raga 1995) or wide-angle outflows 
(e.g. Li $\&$ Shu 1996). Larger scale versions of these structures
have long been known from ground-based NIR imaging (Terebey et al. 1991, 
Kenyon et al. 1993, Lucas \& Roche 1997) and molecular line observations with 
millimeter interferometry (Bontemps et al. 1996, Hogerheidje et al. 1998).  
Hogerheidje (1998) describes the walls
of outflow cavities as regions where outflowing gas interacts with
infalling material in the envelope. Outflow cavities are commonly 
limb-brightened, especially at millimeter wavelengths (e.g. Bontemps et
al. 1996), as seen for DG Tau B and CoKu Tau/1 in the current 
sample, and may be either conical (V-shaped) or paraboloidal in shape.
The object with the smallest clearly defined opening angle for
its V-shaped cavity is DG Tau B which also has the largest mass
of circumstellar material. The single star with the lowest circumstellar
mass (Haro 6-5B) also has the largest opening angle. The two single
sources with intermediate circumstellar masses (IRAS 04016+2610, IRAS
04302+2247) also have morphologies in between these extremes, with conical 
nebulae at short wavelengths and more flattened morphologies at 2 $\micron$.  
This suggests an evolutionary effect by which the cavity walls are widened as
the circumstellar mass decreases.

\par 
The bipolar nebula of Haro 6-5B is unlike the other objects in that
the scattered light is extended parallel, rather than
perpendicular, to the dust lane bisecting
the nebula. In this case, the scattered light appears qualitatively 
similar to models of edge-on disks used to fit
the optical scattered light distribution of HH 30 (Burrows et al. 1996)
and HK Tau/c (Stapelfeldt et al. 1998); however, the inclination is
far enough from edge-on ($\sim$ 10$^{\circ}$; Krist et al. 1998) to permit the 
star to be directly detected in the NIR.  Thus, it appears that the
local circumstellar nebulosity of Haro 6-5B probably traces the
upper and lower surfaces of a flared, optically thick circumstellar disk.
Like HH 30, Haro 6-5B seems to have reached a stage where the scattered 
light distribution is dominated by the disk rather than the envelope, 
but sufficient material remains in the accretion disk to drive an energetic 
stellar jet.

Many YSO outflow ``cavities'' are not completely evacuated of dusty
material. Every object (except possibly Haro 6-5B) that we observe in our 
HST/NICMOS survey has structures within the presumed outflow zone which are 
physically thin but are tens to hundreds of AU in length. In the case of IRAS 04302+2247, the limb brightened cavity walls encompass a plethora
of these filaments. The most spectacular case of a filled outflow cavity is 
IRAS 04248+2612 where the polar regions are centrally brightened by an apparent 
dusty helical outflow channel.  Many of these bright filaments features are 
arcuate, and some form loops elongated in the outflow direction as in the case 
of CoKu Tau/1.  Similar features have been identified in the HST/NICMOS YSO 
observations of Terebey et al. (1999), as well as the ground-based imaging
polarimetric studies of Lucas \& Roche (1997) and Whitney et al. (1997).
The kinematics of these high-surface 
brightness arcs are unknown; however, it is plausible that they are related to 
infalling or outflowing material in the envelope. Repeated observation of such 
features with HST/NICMOS might reveal the dynamics of small scale structures 
within the cavities of YSOs.

\subsection{Effect of Binarity on Circumstellar Morphology}

In the course of our survey, we found two previously unknown binaries with
projected separations of $\sim$ 30 AU. These two sources have the
lowest millimeter continuum fluxes of the YSOs in our sample (cf. Table 1).
This is consistent with the results of
Osterloh \& Beckwith (1995) who found that millimeter continuum
flux was diminished among known binaries with a separation of less than
100 AU. In both cases, we have evidence to suggest a circumbinary disk in the
form of apparent dust lanes relatively well-aligned with the separation vector
of the stellar components. For CoKu Tau/1 the possible dust lane is very thin 
and is only distinguished with difficulty between the curving arcs of
the outflow cavity walls. 

In young close binary systems, theory suggests that the orbital motion of the
stellar components should clear a central hole in any circumbinary disk.
Depending on the eccentricity of the binary orbits, the inner edge of
the circumbinary disk is predicted to be at a distance of 2 - 3 times
the semi-major axis (Artymowicz \& Lubow 1994).
Evidence for central holes in the CoKu Tau/1
and IRAS 04248+2612 systems is lacking for the NICMOS data, since the objects
were selected to be nearly edge-on, and the stellar PSFs make direct
detection of the gap impossible. However, the existence of central
holes in these systems are plausible, given the evidence from other young 
binaries.  Among young stars, the GG Tau circumbinary ring is a spectacular 
face-on example of a disk with a cleared central region (Roddier et al. 1996, 
Koerner, Sargent, \& Beckwith 1993). Despite the central hole, the GG Tau binary
components show ample spectroscopic evidence of accretion, indicating that 
material is bridging the gap and falling onto the stars. The active Herbig-Haro 
jets emanating from both IRAS 04248+2612 and CoKu Tau/1 indicate that accretion 
is continuing in these systems as well.

   Evidence is mounting that disks around individual stars clear from
the inside out, possibly as a result of planet formation processes. It
has long been known that the central 100 AU of the Beta Pictoris
debris disk is depleted of dusty material and this
appears to be true for other debris disks as well (Backman \& Paresce 1993).
Recently, mid-infrared imaging of HR 4796A (Koerner et al. 1998,
Jayawardhana et al. 1998) and submillimeter maps of Epsilon Eridani (Greaves 
et al. 1998) have shown that inner disk holes in the absence of close binary 
companions may be common in younger disks and disks around solar-type stars. 
These stars range in age from about 20 Myr (HR 4796A) to $\sim$ several
hundred million years in age (Epsilon Eridani, Beta Pictoris, etc.). 
However, this disk clearing may occur at a far younger age for close binaries,
dispersing the material required to make planets in the inner $\sim$
100 AU region while retaining a ring of dusty gas in
the outer parts of the circumbinary disk. 

   Despite the apparently diminished disk masses in the two binary
systems, their large scale morphology suggests that both objects
are very young. In CoKu Tau/1, nebulosity stretches for 
many hundreds of AU from the central sources, indicating the presence
of extended circumstellar material in this system. The bipolar
nebula of IRAS 04248+2612 is very elongated in the outflow direction,
appearing as a relatively narrow channel in the surrounding cloud
medium. Both YSOs are in extremely opaque parts of their respective
clouds, without obvious neighbors or background stars. If these
two objects are indeed very young ($\leq$ 1 Myr) sources, they provide further 
evidence that disk evolution may be accelerated in close binaries. 
(Osterloh \& Beckwith 1995, Jensen \& Mathieu 1997, Meyer et al. 1997). 

   It is interesting that the dust lane in IRAS 04248+2612 is not
aligned with the separation vector between the stellar components.
There are at least two possible explanations. The most tantalizing is
that the disk is inclined relative to the orbit plane of the binary. 
Papaloizou \& Terquem (1995) have theoretically modeled
the tidal perturbations of YSO disks by companions on inclined circular
orbits and have suggested that observable warps in the disk are a likely
outcome.  This interpretation has been applied to the warp in the Beta 
Pictoris disk, where an unseen planetary mass companion is presumed to excite 
the disk plane asymmetries (Mouillet et al. 1997).
A more prosaic and likely explanation notes that because the disk is not 
precisely edge-on, the projected binary component separation vector need not 
be in the plane of the disk.  From our vantage point, the orbit of the 
secondary would describe a long and narrow ellipse with minor axis angular 
size equivalent to the tilt of the orbital plane from an edge-on configuration.
Another consequence of this configuration is that the projected
separation of the binary components may be much smaller than their
physical separation. 

\subsection{A Morphological Sequence of Pre-Main Sequence Circumstellar Material?}

   Combining the current NICMOS sample of YSOs with high resolution
observations of young circumstellar disks in the literature, we can begin
to relate high resolution scattered-light morphologies to evolutionary status 
of circumstellar material around solar-type young stars. 
Our goal is to estimate relative placement of the objects in our
small sample in the context of pre-existing evolutionary scenarios.
Presuming that 
millimeter continuum observations are a reasonable measure of circumstellar 
masses, we would order the present objects in the following sequence from 
most massive to least massive: 
DG Tau B ($\sim$ 320 mJy), IRAS 04016+2610 (180 mJy), IRAS 04302+2247
(149 mJy), Haro 6-5b (134 mJy), IRAS 04248+2612 (99 mJy), and
CoKu Tau/1 ($\leq$ 12 mJy). In the objects with greater millimeter
continuum flux such as DG Tau B and IRAS 04016+2610, the NICMOS images
indicate a large amount of material above the midplane in the form
of outflow cavity walls which are optically thick at visual wavelengths,
but become largely transparent at 2 $\mu$m. It is these circumstellar
structures, which can be understood as the interaction zone between the
infalling circumstellar envelope and the polar outflows, which cause
many Class I sources to be very faint in the optical, yet bright at
2 $\mu$m. The extinction in the cavity walls is extremely non-uniform, as
evidenced by the curving swaths of reddening seen above the ``disk''
plane in IRAS 04016+2610 and IRAS 04302+2247. These streamlined
structures could be related to infall as discussed in Terebey et al.
(1999). According to the current paradigm of star formation,
sources with an extensive envelope should be the youngest objects,
and we identify DG Tau B and IRAS 04016+2610 as good examples of
envelope dominated systems. The importance of the envelope component
in the IRAS 04016+2610 system has been highlighted by Hogerheijde et al.
(1997), who found that half or more of the circumstellar mass indicated
by millimeter continuum may originate in a resolved envelope. 
 Envelope systems have been often been called
``optically invisible protostars'' since the extinction in the material above 
the midplane guarantees that the source appears very red even at moderate 
inclinations. 

Among our sample, IRAS 04302+2247 may be at an intermediate stage between
envelope dominated and disk dominated systems. 
Although the NICMOS images show evidence of material above the disk plane, the 
dense matter appears to be confined to near the disk plane. The decreasing 
apparent thickness of its dust lane with increasing wavelength clearly indicates a 
vertical density gradient. The faintness of this object at all wavelengths is a
function of inclination rather than youth; it is seen almost precisely
edge-on. YSOs such as IRAS 04302+2247 are plausibly in a ``big disk/dispersing
envelope'' stage. Probably most of the more active T Tauri stars with
healthy millimeter continuum emission are in a similar stage. A good 
example of a similar object seen at a less favorable inclination is
GM Aurigae (Stapelfeldt et al. 1995b). The disk extends far enough from the 
central stars in this system to be visible as compact circumstellar nebulosity 
even at optical wavelengths. Further analysis of recent millimeter
interferometry for IRAS 04302+2247 should indicate the relative percentage 
of mass in the disk versus the envelope, clarifying the evolutionary
state of this object (Padgett et al. 1999).

The next stage of evolution appears to be
represented by objects such as HH 30 and possibly Haro 6-5B, where the 
scattered light nebulosity adjacent to the star
appears to originate entirely from the surface of the optically
thick disk. The difference between Haro 6-5B and its analogue HH 30
appears to be entirely due to variations in disk mass. It seems plausible
that once the envelope has dispersed, the disk begins to diminish
in mass due to accretion onto the star and planetary system, as well as
associated outflow events. By the time that jet emission has ceased,
edge-on disks will become difficult to see due to the thinness of the
dust lane (Stapelfeldt et al. 1998). We have
yet to see an optically thin disk in optical or NIR wavelengths
around a certifiable solar-type pre-main sequence
star since the IRAS survey was not sensitive enough to detect optically
thin disks at the distance of nearby star-forming regions.
We await WIRE and SIRTF to identify these potentially very
interesting targets for high resolution imaging.

The very lowest circumstellar masses seen for
pre-main sequence stars are in binary systems such as 
IRAS 04248+2612, CoKu Tau/1 and
HK Tau/c (Stapelfeldt et al. 1998), where the effects of the companion
have likely been to clear a hole in a circumbinary disk 
or tidally truncate the circumsecondary disk.  If close
binary systems indeed experience substantially accelerated disk
evolution as suggested by the current study, then this sizable percentage
of YSOs will defy the morphological sequence for single stars, since
their disks may dissipate prior to envelope clearing.
Again, high resolution millimeter observations may clarify
the evolutionary status of these objects.

\subsection{Detailed Morphology Versus IR SED}

In the past decade, our progress in understanding 
star and planet formation has largely depended on interpretation
of young star SEDs from the near-IR to millimeter wavelengths. 
The mid-to-far IR excesses measured by IRAS 
allow us to infer the existence of circumstellar disks and
envelopes around young stellar objects and estimate
their frequency and timescale for dissolution (Strom et al. 1989).
The availability of 15 AU resolution observations of circumstellar structures 
provides a testing ground for models based on SEDs. For example, in the current 
study IRAS 04248+2612 and IRAS 04302+2247 have very similar spectral energy
distributions (Figure 4), especially in the wavelength range covered
by the IRAS satellite. However, comparison of the HST/NICMOS images
for these objects demonstrate that the detailed morphology of these
sources is quite distinct from each other, as discussed in Sections
3.2 and 3.3. In particular, the IRAS fluxes do not provide any
indication of binarity for IRAS 04248+2612. Since the IRAS bands
are sensitive to circumstellar material at $\leq$ 1 AU - 100 AU,
it would seem that small circumstellar disks are probably still
present in the close binary system.  The only hint of binarity found
in the SED is the lower 1 mm continuum flux for
IRAS 04248+2612 (Osterloh \& Beckwith 1995).  The inclination of
IRAS 04302+2247 is suggested by its IRAS non-detection at
12 microns. Radiation transfer models of edge-on disks suggest
that they are viewed primarily in scattered light,
and are thus exceptionally faint, out to wavelengths
exceeding 25 $\mu$m (Sonnhalter et al. 1995). ISO observations
of the edge-on disk system HH 30 appear to confirm these predictions
for that object (Stapelfeldt \& Moneti 1998).
The significance of factors such as source multiplicity, disk extent, 
orientation, and gaps, dynamical structures in the polar outflow zones,
and interstellar environment can only be determined by imaging at scales
of smaller than tens of AU. Modelling of high-resolution optical and NIR
images complements information derived from SED modelling in providing an 
accurate depiction of the circumstellar environment of YSOs. The 
extremely high sensitivity and improved resolution of upcoming mid-to-far
IR space telescopes such as WIRE and SIRTF will extend the IRAS results
to more evolved systems and identify potential new targets for
high-resolution imaging.  

\section{Conclusions}

We observed six young stellar objects at 1.1 $\micron$ - 2
$\micron$ with HST/NICMOS. Images with 15 AU resolution were
successfully obtained for young stellar objects systems spanning a 
range of circumstellar 
masses as derived from millimeter continuum emission. The
near-infrared morphology reveal conical to parabolic bipolar reflection
nebulae crossed by dark lanes. We identify the dark lanes as
circumstellar disks of dimension 500 AU - 1000 AU seen at near
edge-on inclinations. Millimeter interferometry available for
some of the sources provides evidence of rotational motion for
material in these disks. We identify the bipolar reflection
nebulae as infalling envelopes illuminated by the central stars,
and, in some cases, the top and bottom surfaces of optically thick
circumstellar disks. The limb-brightened cavities noted in several
sources most likely represents the boundary between the infalling
envelope and accretion-driven outflow. If so, the opening angles
of the cavities are in contrast to the narrowly collimated jets
observed within the cavities for some sources. Evolutionary
effects are suggested by the increase in cavity opening angle
as disk mass (traced by millimeter continuum emission) decreases.
In addition, although two sources found in our study to be close
binaries have the smallest disk masses, their extended morphology
and IRAS SEDs closely resembles other envelope-dominated YSOs. This
suggests accelerated disk evolution for close binary stars.

\acknowledgements{The authors acknowledge contributions by S.
Kenyon, as well as helpful comments by an anonymous referee. 
Deborah Padgett and Wolfgang Brandner gratefully acknowledge 
support from the NASA/WIRE project and STScI, as well as the tireless
efforts of the NICMOS instrument team in making these observations possible. 
Support for this work was provided by NASA through grant number
STScI GO-7418.01-96A from the Space Telescope Science Institute,
which is operated by the Association of Universities for Research in
Astronomy, Incorporated, under NASA contract NAS5-26555}

\begin{deluxetable}{lrrrcrc}
\tablecaption{Taurus Young Stellar Object Sample.\label{tbl-1}}
\footnotesize
\tablehead{\colhead{Target}
&\colhead{RA (J2000)} & \colhead{Dec (J2000)} & \colhead{K} &
\colhead{Outflow} & \colhead{$\theta_{of}$ ($^{\circ}$)} &
\colhead{MM Continuum Flux (mJy)}}
\startdata
IRAS 04016+2610 & 04 04 42.91 & +26 19 02.3 & 9\fm33 & optical,mm
& 180 \tablenotemark{1} & 180 \tablenotemark{2} \\
CoKu Tau/1 & 04 18 51.50 & +28 20 28.0 & 10\fm85 & optical
& 225 \tablenotemark{3} & $\leq$ 12 \tablenotemark{4} \\
Haro 6-5B & 04 22 00.69 & +26 57 33.3 & 9\fm89 & optical
& 60 \tablenotemark{5} & 134 \tablenotemark{6}
\\
DG Tau B & 04 27 02.55 & +26 05 30.9 & 11\fm52 & optical,mm
& 122 \tablenotemark{7} & 310 \tablenotemark{8} \\
IRAS 04248+2612 & 04 27 57.32 & +26 19 19.0 & 10\fm65 & optical
& 172 \tablenotemark{9} & 99 \tablenotemark{10} \\
IRAS 04302+2247 & 04 33 16.45 & +22 53 20.7 & 11\fm07 & optical?,mm?
& 160 \tablenotemark{9} & 149 \tablenotemark{10} \\
\enddata
\tablenotetext{1}{Hogerheijde et al. 1998}
\tablenotetext{2}{1.1 mm; Ohashi et al. 1991}
\tablenotetext{3}{Strom et al. 1986}
\tablenotetext{4}{1.3 mm; Osterloh \& Beckwith 1995}
\tablenotetext{5}{Eisl\"offel \& Mundt 1998}
\tablenotetext{6}{1.3 mm; Dutrey et al. 1996}
\tablenotetext{7}{Mitchell et al. 1997}
\tablenotetext{8}{1.3 mm; Stapelfeldt et al. 1999}
\tablenotetext{9}{Gomez et al. 1997}
\tablenotetext{10}{1.1 mm; Moriarty-Schieven et al. 1995}
\end{deluxetable}

\begin{deluxetable}{lrcccc}
\tablecaption{Observing log for GO 7418.\label{tbl-2}}
\footnotesize
\tablehead{\colhead{Target}
&\colhead{Date}    & \colhead{F110W} & \colhead{F160W} & \colhead{F187W}& \colhead{F205W}}
\startdata
IRAS 04016+2610 & Dec 20, 97& 768s & 768s & 512s & 256s\\
CoKu Tau/1 & Aug 19, 97& 768s & 512s & 512s & 512s\\
Haro 6-5B & Oct 29, 97& 768s & 512s & 512s & 512s\\
DG Tau B & Dec 21, 97& 768s & 512s & 512s & 512s\\
IRAS 04248+2612 & Sep 23, 97& 1024s & 512s & 512s & 256s\\
IRAS 04302+2247 & Aug 19, 97& 768s & 512s & 512s & 512s\\
\enddata
\end{deluxetable}

\begin{deluxetable}{lcccc}
\tablecaption{Flux values (mJy) (3\farcs8 aperture).\label{tbl-3}}
\footnotesize
\tablehead{\colhead{Target}
& \colhead{F110W} & \colhead{F160W} & \colhead{F187W}& \colhead{F205W}}
\startdata
DG Tau B & 0.59$\pm$0.03  & 4.75$\pm$0.10  & 8.50$\pm$0.17  & 10.7$\pm$0.1 \\
IRAS 04016+2610 & 0.30$\pm$0.01 & 9.58$\pm$0.14  & 21.4$\pm$0.3  &41.6$\pm$0.2\\
IRAS 04302+2247 & 0.64$\pm$0.02 & 6.61$\pm$0.12  & 10.2$\pm$0.2  &11.3$\pm$0.1\\
Haro 6-5B & 0.77$\pm$0.03  & 5.79$\pm$0.11 & 10.9$\pm$0.2 & 14.6$\pm$0.1  \\
IRAS 04248+2612 & 6.13$\pm$0.10  & 20.9$\pm$0.2  & 21.3$\pm$0.3 & 20.5$\pm$0.2\\
CoKu Tau/1 & 6.25$\pm$0.09 & 26.8$\pm$0.2& 30.2$\pm$0.3& 29.1$\pm$0.2 \\
\enddata
\end{deluxetable}

\begin{deluxetable}{rcccc}
\tablecaption{Point Source Fluxes (mJy). \label{tbl-4}}
\footnotesize
\tablehead{\colhead{Target} &\colhead{F110W} & \colhead{F160W} & \colhead{F187W}& \colhead{F205W} }
\startdata
DG Tau B & 0.033$\pm$0.006 & 0.48$\pm$0.04 & 1.25$\pm$0.08 & 2.75$\pm$0.07\\
IRAS 04016+2612 & 0.021$\pm$0.004& 1.29$\pm$0.04& 7.88$\pm$0.15&27.4$\pm$0.17 \\
IRAS 04302+2247 & $<$0.00056  & $<$0.0023  & $<$0.0048  & $<$0.014 \\
Haro 6-5B & 0.39$\pm$0.03 & 2.50$\pm$0.08  & 5.59$\pm$0.17  & 8.64$\pm$0.13 \\
IRAS 04248+2612 A \tablenotemark{1}& 0.85$\pm$0.06&3.50$\pm$0.13 & 4.70$\pm$0.19  & 5.80$\pm$0.12 \\
IRAS 04248+2612 B \tablenotemark{1}&(0.76$\pm$0.06)&2.90$\pm$0.11& 3.70$\pm$0.17  & 4.00$\pm$0.10  \\
CoKu Tau/1 A \tablenotemark{2}& 2.78$\pm$0.07  & 11.1$\pm$0.2& 13.3$\pm$0.24  & 13.2$\pm$0.1  \\
CoKu Tau/1 B \tablenotemark{2}& (0.56$\pm$0.03)  & 3.3$\pm$0.1  & 4.7$\pm$0.15  & 5.1$\pm$0.1  \\
\enddata

\tablenotetext{1}{The eastern component is designated A.  The measured
binary separation is 0\farcs160$\pm$0\farcs002 at PA
$266^{\circ}\pm0.5^{\circ}$.}
\tablenotetext{2}{The western component is designated A.  The measured
binary separation is 0\farcs240$\pm$0\farcs002 at PA
$111^{\circ}\pm1.4^{\circ}$.}

\end{deluxetable}

\begin{deluxetable}{rcccc}
\tablecaption{Morphological Parameters of Dust Lanes\label{tbl-5}}
\footnotesize
\tablehead{\colhead{Target} & \colhead{PA ($^{\circ}$)} &
\colhead{Major axis (AU)} & \colhead{Minor Axis (AU)} &
\colhead{PSF Visibility}}
\startdata
DG Tau B & 32 & 600 & 210 & F187W+ \\
IRAS 04016+2610 & 85 & 550? & 240 & F160W+ \\
IRAS 04302+2247 & 175 & 1000 & 340 (110W) - 250 (205W) & None \\
Haro 6-5B & 147 & 600 & 150 & All filters \\
IRAS 04248+2612 & 82 & 600  & 220 - 126 & All filters \\
CoKu Tau/1 & 120 & 600 & 50 - 100 & F110W+, F160W+ \\
\enddata
\end{deluxetable}

\begin{deluxetable}{lcccc}
\tablecaption{Morphological Parameters of Reflection Nebulae \label{tbl-6}}
\footnotesize
\tablehead{\colhead{Target} &\colhead{$\theta_{sa}$ ($^\circ$)}
& \colhead{Major axis (AU)}
& \colhead{Minor Axis (AU)} & \colhead{Opening Angle ($^\circ$)}}
\startdata
DG Tau B & 122 & 550 & 350  & 83(e), 45(w) \\
IRAS 04016+2610 & 175 & 1100 & 850 & 65(1$\micron$)
-160(2$\micron$) \\
Haro 6-5B & 57 & 560 & 210 & 160 \\
IRAS 04302+2247 & 85 & 850 & 400 & 60 \\
IRAS 04248+2612 & 170 & 950  & 560 & 60 \\
CoKu Tau/1 & 210 & 900 & 550 & 92 \\
\enddata
\end{deluxetable}

%
%
\begin{sidefigure}
\psfig{file=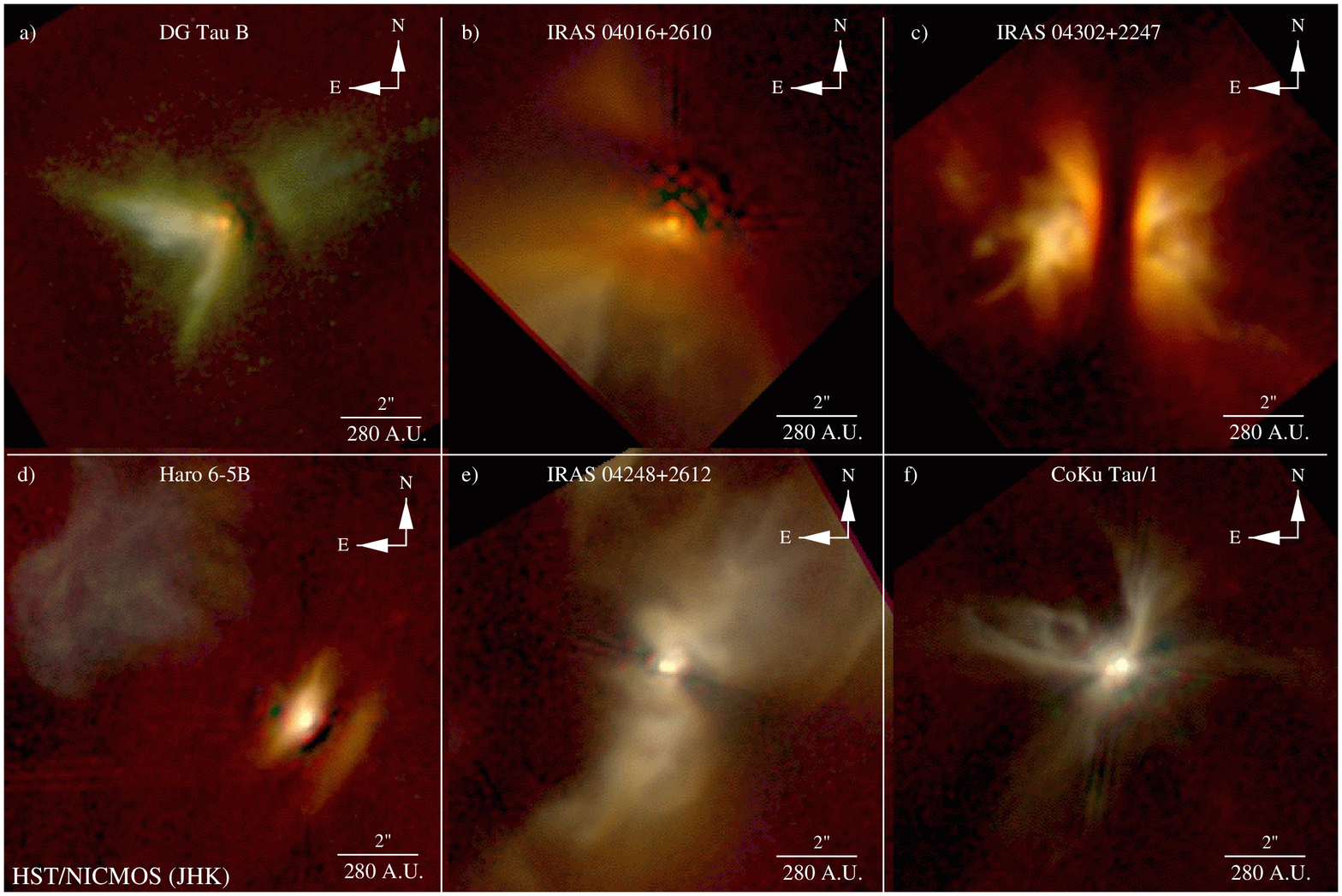,width=22cm,angle=270}
\figcaption{HST/NICMOS images of Taurus young stellar objects, arranged
in order of decreasing circumstellar mass. These are pseudo-true color 
composites of NICMOS F110W (1.1 $\micron$), F160W (1.6 $\micron$), and
F205W (2.05 $\micron$) broad-band observations. Each image was deconvolved
using theoretical point-spread functions, resulting in a factor of 3
reduction in extended PSF features. 
Note that objects e) and f) are subarcsecond binaries.
North is up in all images. \label{f1}} 
\end{sidefigure}
%
%
%
\begin{figure}
\psfig{file=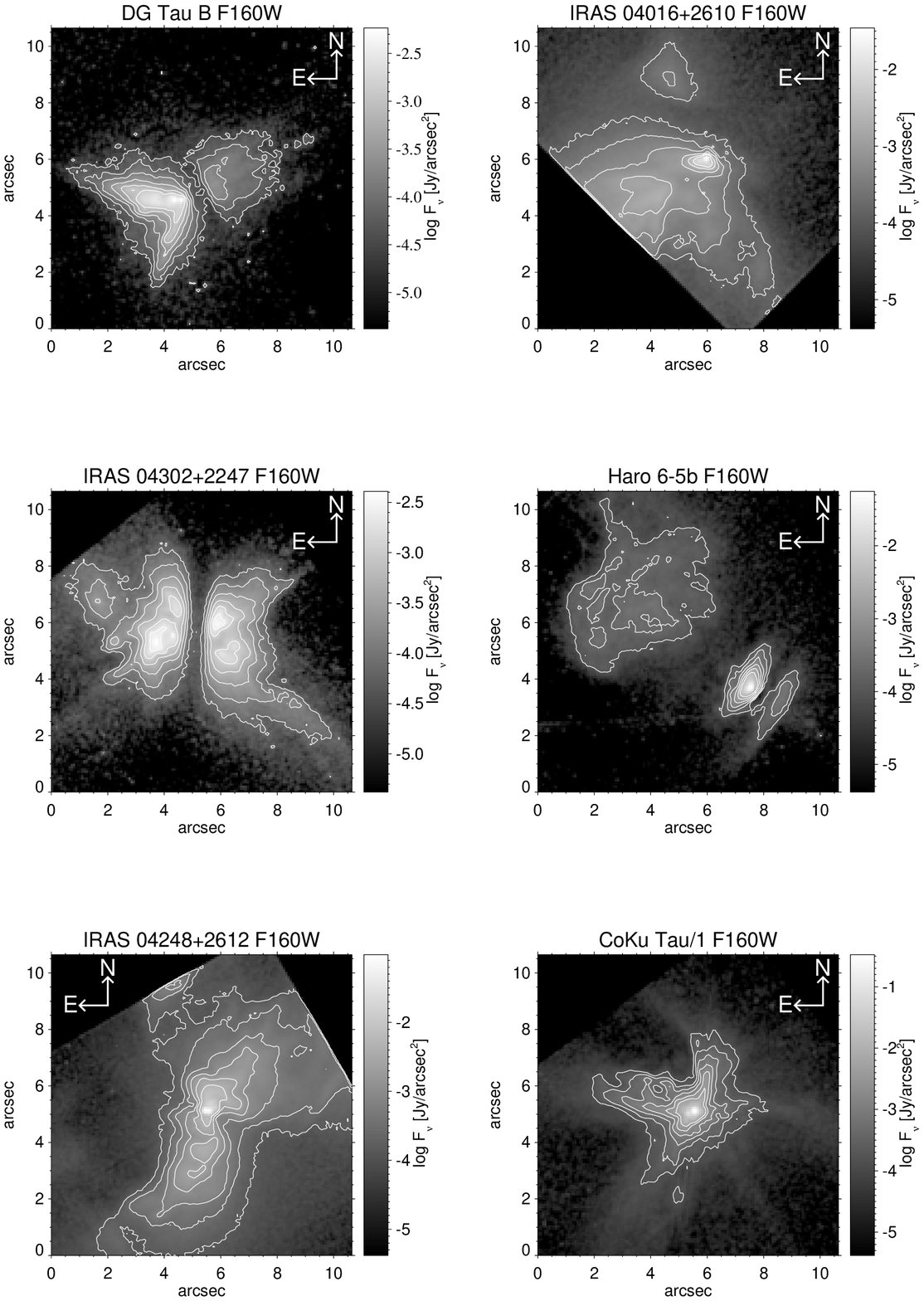,width=14cm}
\figcaption{Calibrated HST/NICMOS F160W (1.6 $\micron$) broad-band
surface photometry for Taurus young stellar objects. As in
Figure 1, the images have been deconvolved to
a 0\farcs115 resolution. Contour levels (from left to right, and from
top to bottom) start at 0.04 mJy, 0.3 mJy, 0.3 mJy, 0.8 mJy, 2.4 mJy,
and 4.7 mJy, respectively,
 and then double in flux with each subsequent contour level.
\label{f2}} 
\end{figure}
%
%
%
\begin{figure}
\psfig{file=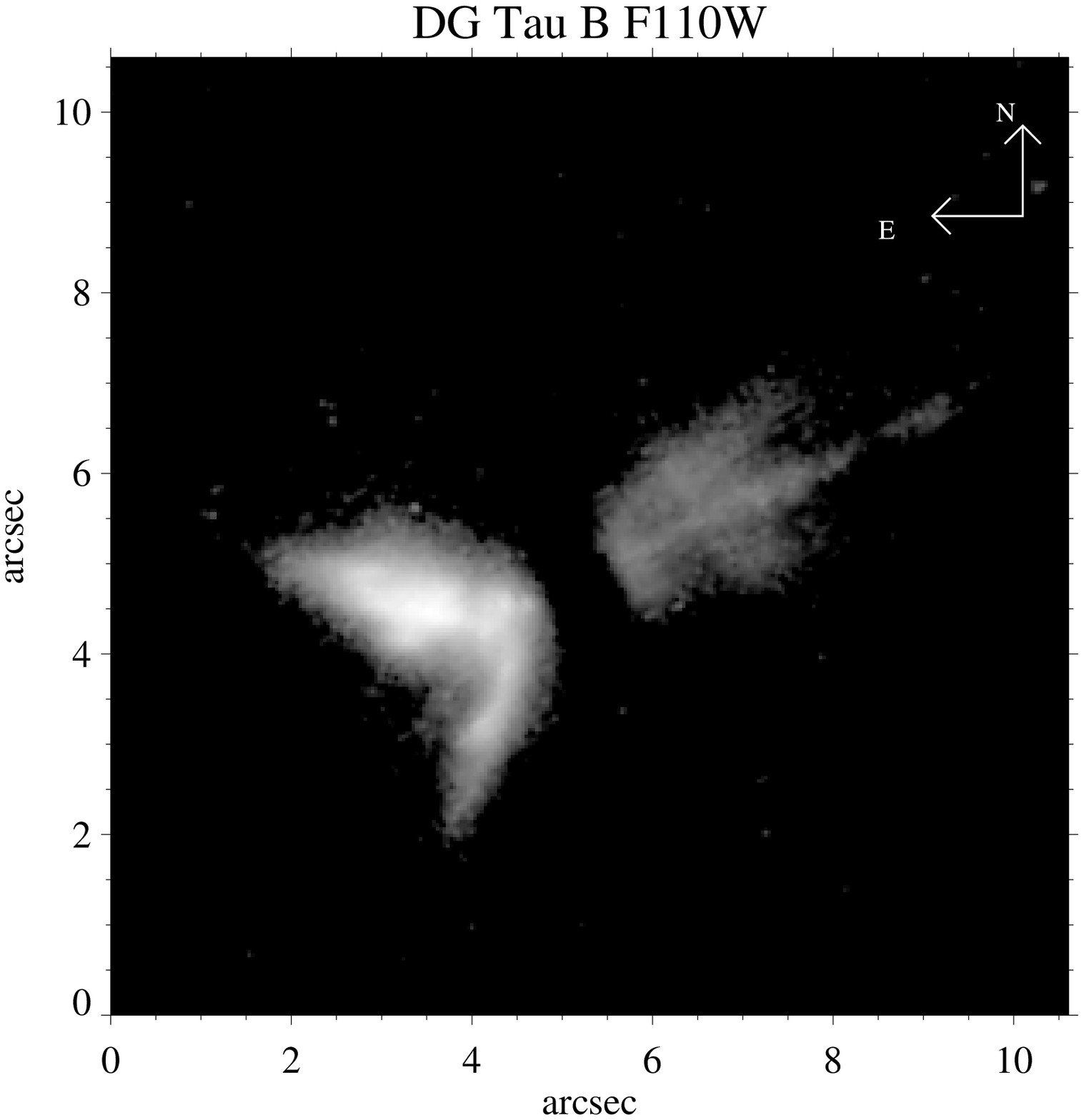,width=16cm,angle=90}
\figcaption{HST/NICMOS F110W (1.1 $\micron$ broad-band) image of DG Tauri B.
Note the linear jet emerging opposite the V-shaped reflection nebula. 
Emission from jet knots is probably from [Fe II] 1.26 $\micron$. \label{f3}}
\end{figure}
%
%
%
\begin{figure}
\psfig{file=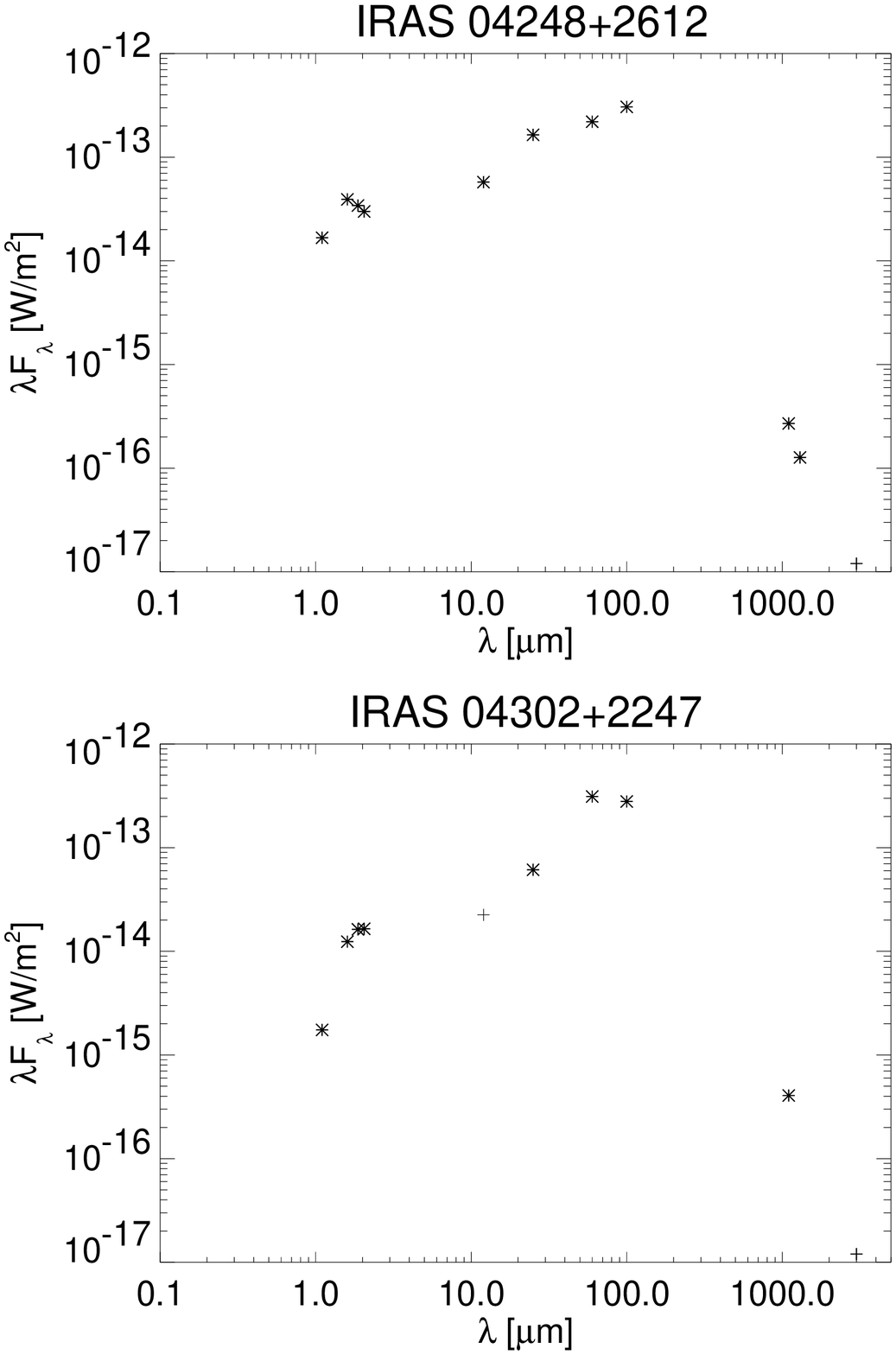,width=13cm}
\figcaption{Comparison of spectral energy distributions for IRAS 04302+2247
(edge-on Class I YSO) and IRAS 04248+2612 (subarcsecond binary Class I YSO).
Asterisks represent measured values, and pluses represent upper limits.
Near-infrared points are from the current study, and IRAS fluxes are taken
from Beichmann et al. (1992). Millimeter flux densities are from Saraceno
et al. (1996), Ohashi et al. (1996), Moriarty-Schieven et al. (1996), 
and Hogerheijde et al. (1997).  Despite the morphological differences between 
these two objects, the SEDs are very similar.  In addition, I04248 has a lower 
1 mm flux than I04302, possibly due to clearing of the central portion of the 
circumbinary disk by the orbital motion of the stars.  \label{f4}} 
\end{figure}

\end{document}